\documentclass[a4paper,english,aps,floats,twocolumn,showpacs,nofootinbib]{revtex4}
\usepackage{pslatex}
\usepackage[T1]{fontenc}
\usepackage{graphicx}
\usepackage{epsfig}

\usepackage{calc}
\usepackage{ifthen}
\usepackage{subfigure}

\newcommand{\bea}{\begin{eqnarray}}
\newcommand{\eea}{\end{eqnarray}}

\newcommand{\nc}{\newcommand}
\nc{\renc}{\renewcommand}
\nc{\eqs}[2]{\mbox{Eqs.~(\ref{#1},\,\ref{#2})}}
\nc{\eq}[1]{\mbox{Eq.~(\ref{#1})}}
\nc{\figs}[2]{\mbox{Figs.~(\ref{#1},\,\ref{#2})}}
\nc{\fig}[1]{\mbox{Fig~.(\ref{#1})}}
\nc{\be}[1]{\begin{equation} \mbox{$\label{#1}$}}
\nc{\ee}{\vspace{0.1cm}\end{equation}}

\newcommand{\bean}{\begin{eqnarray*}}
\newcommand{\eean}{\end{eqnarray*}}

\def\GeV{{\rm \ GeV}}

\begin{document}
\title{Higgs Inflation and Naturalness}
\author{Rose N. Lerner}
\email{r.lerner@lancaster.ac.uk}
\author{John McDonald}
\email{j.mcdonald@lancaster.ac.uk}
\affiliation{Cosmology and Astroparticle Physics Group, University of
Lancaster, Lancaster LA1 4YB, UK}
\begin{abstract}

  Inflation based on scalar fields which are non-minimally coupled to gravity has been proposed as a way to unify inflation with weak-scale physics, with the inflaton being identified with the Higgs boson or other weak-scale scalar particle. These models require a large non-minimal coupling $\xi \sim 10^{4}$ to have agreement with the observed density perturbations. However, it has been suggested that such models are unnatural, due to an apparent breakdown of the calculation of Higgs-Higgs scattering via graviton exchange in the Jordan frame. Here we argue that Higgs inflation models are in fact  natural and that the breakdown does not imply new physics due to 
strong-coupling effects or unitarity breakdown, but simply a failure of perturbation theory in the Jordan frame as a calculational method. This can be understood by noting that the model is completely consistent when analysed in the Einstein frame and that scattering rates in the two frames are equal by the Equivalence Theorem for non-linear field redefinitions.

 \end{abstract}
\maketitle

\section{Introduction}

There has recently been a revival of the idea \cite{salopek} that non-minimally coupled scalar fields can account for the flat potential
required to account for inflation without very small scalar couplings \cite{bs1,bgs,bms,bs2,barv1,barv2,desimone,gb,clark,rl}.
This has been used to suggest that the Higgs itself could be the inflaton \cite{bs1}, as part of a minimal approach to particle physics and cosmology in which there are no other mass scales between the weak scale and the Planck scale \cite{shapmin}. This has been extended to incorporate a minimal thermal WIMP dark matter candidate, namely a gauge singlet scalar, \cite{clark,rl}, with
inflation along the Higgs direction being studied in \cite{clark} and
inflation along the singlet direction ("S-inflation") being studied in \cite{rl}. 
The Higgs inflation model has recently been extended to include see-saw neutrino masses \cite{shafi} and a SUSY version has been proposed \cite{tj}. (See \cite{others} for related studies.)

  A feature of these models is that the non-minimal coupling of the Higgs scalar to the Ricci scalar, $\xi$, must be large in order to account for the observed curvature perturbation, $\xi \sim 10^4$. This has led some authors to question the consistency of the model at large energy. Specifically, in \cite{barbon}, it was noted that Higgs-Higgs scattering via s-channel graviton exchange becomes strongly-coupled
at a scale $\Lambda \approx M_{P}/\xi$ (where $M_{P} = 2.4 \times 10^{18} \GeV$ is the reduced Planck scale). This is much smaller than the Higgs expectation value at $N$ e-foldings during inflation, $h \approx \sqrt{N} M_{P}/\sqrt{\xi}$. In \cite{burgess}, using a general loop counting method in the context of effective field theory, it was proposed that unitarity breaks down at $\Lambda$. In both cases, $\Lambda$ is interpreted as a new physics scale, either due to strong-coupling effects or due to the unitarity-preserving completion of the model, which implies new operators in the scalar potential, scaled by powers of $\Lambda$.
Such operators would spoil the flatness of the inflaton potential, making inflation unnatural. These objections have been cited as strong arguments against Higgs-type inflation (e.g. \cite{tamvakis}).

In this letter we revisit the arguments against Higgs-type inflation. Our focus will be the discussion of \cite{barbon}, but we will also comment on \cite{burgess}. Our main point is that the arguments given in \cite{barbon} do not demonstrate the need for new physics at the scale $\Lambda$. They only show that perturbation theory for Higgs scattering via graviton exchange breaks down at $E \sim \Lambda$.
In particular, when the model is analysed in the Einstein frame, there is no breakdown of the theory at $E$
 or $\chi$  $ \gtrsim \Lambda$. At $\chi \gtrsim \Lambda$, where $\chi$ is the canonically normalized scalar field in the Einstein frame, all that happens is a smooth change in the behaviour of the $\chi$ scalar potential from $\chi^4$ to $\chi^2$. In addition, the Equivalence Theorem for non-linear field redefinitions \cite{et,flan} states that the S-matrix for scattering processes is invariant under non-linear redefinitions, which implies that the Higgs scattering rate should be the same in the Jordan and Einstein frame. The scalar in the Einstein frame is minimally coupled to gravity, perturbative and unitary throughout, with no increase in the $\chi$-$\chi$ scattering cross-section expected at $E \sim \Lambda$.
Therefore no new physics is implied by the $\chi$-$\chi$ scattering cross-section in the Einstein frame. Therefore, by the Equivalence Theorem, no new physics is necessary to understand the Higgs-Higgs cross-section in the Jordan frame.  It follows that the breakdown of perturbation theory in the Jordan frame at $E \sim \Lambda$ is simply a failure of the perturbation method in the Jordan frame, not a sign of new physics leading to the generation of terms in the Higgs potential scaled by $\Lambda$. In a nutshell, since the model is completely consistent and natural in the Einstein frame, it must be completely consistent and natural in the Jordan frame.

The remainder of this letter proceeds as follows. In Sec. II, we review Higgs inflation, concentrating on the transformation between, and equivalence of, the Jordan and Einstein frames. In Sec. III, we present our arguments for the naturalness of the theory and address the comments of \cite{barbon,burgess} before concluding in Sec. IV.

 \section{Jordan and Einstein Frame in Higgs Inflation}

We first review the Higgs inflation model and the transformation from the Jordan to the Einstein frame. The Jordan frame metric, $g_{\mu\nu}$, is defined to relate coordinates to physical space and time. The Higgs inflation model is first formulated in this frame as the Standard Model with a non-minimal coupling of the Higgs to the Ricci scalar \cite{bs1}:

\bea
\label{Jaction}
 S_J & = & \int \sqrt{-\!g}\,d^4\! x \Big({\cal L}_{\overline{SM}} + \left(\partial _\mu H\right)^{\dagger}\left(\partial^\mu H\right)  \nonumber \\
 & &  - \frac{M^2R}{2} - \xi H^{\dagger}H R  - V(H^{\dagger}H)\Big)
~,\eea
where ${\cal L}_{\overline{SM}}$ represents the Standard Model Lagrangian excluding the Higgs fields and
\be{e1}
\label{Jpot}
V(H^{\dagger}H)  = \lambda \left(\left(H^{\dagger}H\right) - \frac{v^2}{2}\right)^2  ~.\ee
From now we consider only the physical Higgs field $h$, where $H~=~\frac{1}{\sqrt{2}} \left(\begin{array}{c} 0 \\ h + v \end{array}\right) $, $v = 246 \GeV$ is the Higgs expectation value and $h$ is real.

The Einstein frame action is obtained by first performing a
conformal rescaling of the metric
\be{2} \tilde{g}_{\mu\nu} = \Omega ^2 g_{\mu\nu} ~,\ee
where
\be{3}\label{omegaeq} \Omega ^2 = 1 + \frac{\xi h^2}{M_P^2} ~.\ee
In terms of this metric the action becomes
$$ S_{E} = \int d^{4} x  \sqrt{-\tilde{g}} \left( - \frac{M^{2}}{2}\tilde{R} + \frac{1}{2 \Omega^{4}} \left( \Omega^{2}
 + \frac{6 \xi^{2}h^{2}}{M^{2}} \right)\tilde{\partial}_{\mu}h \tilde{\partial}^{\mu}h \right.$$
\be{e4} \left. - \frac{\lambda}{4 \Omega^{4}} \left(h^2 - v^2\right)^{2}  \right) ~,\ee
where $\tilde{R}$ is the Ricci scalar with respect to $\tilde{g}_{\mu \nu}$ and $\tilde{\partial}_{\mu}$ indicates derivatives contracted with $\tilde{g}_{\mu\nu}$.
This can be written in terms of a canonically normalised scalar $\chi$ via the redefinition:
\be{e5} \frac{d\chi}{dh} = \sqrt{\frac{\Omega ^2 + 6 \xi^2h^2/M_P^2}{\Omega ^4}} ~.\ee
Then
\be{e6}
S_E = \int d^4x\sqrt{-\tilde{g}}\Big( - \frac{M_P^2\tilde{R}}{2} + \frac{1}{2}\tilde{\partial} _\mu \chi \tilde{\partial}^{\mu} \chi - U(\chi)\Big)
~,\ee
where
\be{e7} U(\chi) = \frac{1}{\Omega^4}V(h) \equiv \frac{1}{\Omega^4} \left( \frac{\lambda}{4}(h^2 - v^2)^2 \right) ~.\ee

   There are three distinct ranges of $h$, which correspond to different behaviours of the $\chi$ scalar potential in the Einstein frame:-

\noindent \noindent ${\bf h < M_{P}/\sqrt{6} \xi }$:  In this case $d \chi/d h \approx 1$, $\Omega^2 \approx 1$ and $\chi \approx h$. Therefore
\be{e8} U(\chi) \approx \frac{\lambda}{4} \chi^{4}   ~.\ee

\noindent \noindent ${\bf M_{P}/\sqrt{6}\xi < h < M_{P}/\sqrt{\xi}}$:  In this case $d \chi/d h \approx \sqrt{6} \xi h/M_{P}$, $\Omega^2 \approx 1$ and
\be{e8a} \chi \approx \frac{\sqrt{6} \xi h^2}{2 M_{P}}   ~.\ee
Therefore
\be{e9} U(\chi) \approx \frac{\lambda}{6} \frac{M_{P}^{2} \chi^{2}}{\xi^{2}}  ~.\ee

\noindent \noindent ${\bf M_{P}/\sqrt{\xi} < h}$:  In this case $d \chi/d h \approx \sqrt{6} M_{P}/h$, $\Omega^2 \approx \xi h^2/M_{P}^2$ and
\be{e10} \chi \approx \sqrt{6} M_{P} \ln{\left( \frac{\sqrt{\xi}h}{M_{P}} \right)}  ~.\ee
Therefore
\be{e11} U(\chi) \approx \frac{\lambda}{4} \frac{M_{P}^{4}}{\xi^{2}} \left( 1 + \exp\left(-\frac{2 \chi}{\sqrt{6} M_{P}}
\right)\right)^{-2} ~.\ee
This potential becomes very flat at large values of $\chi$ ($\chi \gg \sqrt{6}M_{P}/2$) and therefore can drive inflation.

\subsection{Calculations in the Einstein frame}

Once we are in the Einstein frame, we can do calculations without any reference to the Jordan frame, as if we had originally formulated the theory in the Einstein frame. This means that slow-roll inflation can be entirely studied in the Einstein frame by assuming a FRW metric in the Einstein frame. To do this we need to make a redefinition (coordinate transformation) of $t$. The Einstein frame line element obtained by rescaling the FRW metric in the Jordan frame is
\be{e12} d \tilde{s}^2 =  \Omega^2 dt^2 - \Omega^2 a^{2}(t) d \bf{x}^2   ~,\ee
where $a(t)$ is the scale factor in the Jordan frame. Rescaling
$dt \rightarrow d\tilde{t} = \Omega dt$ and defining $\tilde{a} = \Omega a$ then gives a FRW metric in the Einstein frame
\be{e13} d \tilde{s}^2 =  d\tilde{t}^2 - \tilde{a}^{2}(t) d \bf{x}^2   ~.\ee
After this coordinate transformation, the Einstein frame action in terms of the metric \eq{e13} has the same form as \eq{e6}. The equations describing inflation in the Einstein frame are then the conventional slow-roll inflation equations, and the analysis of quantum fluctuations of the curvature follows the standard procedure.

A key point in the analysis of inflation in the Einstein frame is that the observable quantities from inflation are associated with perturbations when they re-enter the horizon at late times.  At this time, $h$ (or $\chi$) is small compared with $M_{P}/\sqrt{\xi}$ and $\Omega \rightarrow 1$. Therefore the metric and the curvature perturbation in the Jordan and Einstein frames at late times are identical, although during inflation the curvature perturbation in the Einstein frame is not the physical curvature perturbation. Therefore one can use the standard slow-roll inflation machinery in the Einstein frame to compute the spectral index and other inflation observables in the Jordan frame\footnote{However, as shown in the Appendix, the definition of the number of e-foldings of inflation is different in the Einstein and Jordan frames.}. This argument is true provided that we are able to correctly calculate the curvature perturbation by working in the Einstein frame. It is not obvious that quantizing metric and scalar field perturbations in the Jordan and Einstein frames are equivalent (as the metric and scalar field degrees of freedom are different in the two frames). However, it has been shown that the curvature perturbation computed in the Jordan and Einstein frames are equal \cite{equal}.

\section{Naturalness of Higgs inflation}
In the previous section, we showed that the Higgs inflation model can be entirely analysed in the Einstein frame, where it behaves as a routine minimally coupled inflation model which can be studied using the standard machinery of slow-roll inflation. We now discuss the naturalness of the model. By "natural" we mean that the inflaton potential does not receive corrections scaled by powers of a new physics scale $\Lambda$, which would spoil the flatness of the potential. Since $\chi$ is minimally coupled, there are no problems related to $\chi$-$\chi$ scattering via graviton exchange. The scalar potential, although non-renormalizable, has no terms resulting in the generation of large corrections to the potential scaled by $\Lambda= M_p / \xi$. 
In fact, the $\chi$ potential becomes progressively flatter as $\chi$ increases, corresponding to increasingly light quanta which are more weakly interacting (on perturbing about the zero-mode $\chi$ field). The only change at $\chi \approx \Lambda $ is a smooth transition from an approximately $\chi^4$ potential to a $\chi^2$ potential i.e. from an interacting to a non-interacting scalar field. There is nothing in the Einstein frame analysis which indicates a breakdown of perturbation theory or unitarity or the need for new physics.

Therefore Higgs inflation is a completely consistent, well-behaved theory when considered in the Einstein frame, with no
 new physics necessary at $E$ or $\chi \approx \Lambda$.

We next address the specific concerns of \cite{barbon,burgess} in turn.

\subsection{Higgs-Higgs scattering}

    The apparent problems of Higgs inflation come from consideration of Higgs-Higgs scattering via s-channel graviton exchange in the Jordan frame. Note that in this case we are considering scattering of two Higgs bosons in the vacuum ($\langle h\rangle =  v$) and not the Higgs field during inflation.

    The Jordan frame metric is expanded about flat spacetime $\eta_{\mu \nu}$
\be{e21}  g_{\mu \nu} = \eta_{\mu \nu} + M_{P}^{-1} \gamma_{\mu \nu}   ~.\ee
Then
\be{e22}   R = M_{P}^{-1}[\partial_{\mu}\partial^{\mu} \gamma_{\nu}^{\nu} -  \partial_{\mu}\partial_{\nu} \gamma^{\mu \nu}] + O(\gamma^2)     ~.\ee
Therefore
\be{e23} \frac{\xi h^{2} R }{2} \rightarrow \frac{\xi h^2 }{2 M_{P}} [\partial_{\mu}\partial^{\mu} \gamma_{\nu}^{\nu} -  \partial_{\mu}\partial_{\nu} \gamma^{\mu \nu}]   ~.\ee
The first term
\be{e24} \frac{\xi h^2 }{2 M_{P}} \eta^{\mu \nu}\partial^{2} \gamma_{\mu \nu}       ~,\ee
gives a $hh\gamma$ vertex ($\gamma \equiv$ graviton). For a graviton propagator with energy $E$, the effective dimensionless coupling of this vertex is of order $\xi E/M_{P} \equiv E/\Lambda$. Therefore the perturbative calculation of Higgs-Higgs scattering via s-channel graviton exchange breaks down when $E \gtrsim \Lambda$.

In \cite{barbon} this is interpreted as the onset of a new physics regime and it is proposed that new operators scaled by powers of $\Lambda$ must therefore be introduced. However, the breakdown of perturbation theory does not automatically imply new physics. It could simply mean that perturbation theory cannot be applied at $E \gtrsim \Lambda$, even though the full (non-perturbative) Higgs-Higgs scattering cross-section may be a smooth function of $E$ with no new physics at $E \approx \Lambda$. This is supported by the well-behaved nature of the model in the Einstein frame.  Since Higgs-Higgs scattering is occurring in the vacuum with flat spacetime, if we consider the asymptotic initial state at $t \rightarrow -\infty$, then $h$ and $\chi$ are indistinguishable since $\Omega = 1$ in the vacuum. We can then consider $\chi$-$\chi$ scattering in the Einstein frame and obtain the asymptotic final state at $t \rightarrow + \infty$, where again $h$ and $\chi$ are indistinguishable. Since in the Einstein frame there is no strong coupling or unitarity breakdown (the potential becomes more weakly interacting at large $\chi$), the scattering cross-section can be computed perturbatively without the need for new physics.
According to the Equivalence Theorem of models related by non-linear transformations of the fields \cite{et}, in particular by conformal transformations (which includes the Jordan and Einstein frame formulations of Higgs inflation) \cite{flan}, the S-matrix for scattering processes is invariant under such transformations. Therefore the Einstein frame $\chi$-$\chi$ cross-section should be the same as the Jordan frame $h$-$h$ cross-section. This means that no new physics is necessary to understand the $h$-$h$ cross-section in the Jordan frame. If it were necessary to introduce new physics in Jordan frame in order to understand the $h$-$h$ cross-section (generating new terms in the Higgs potential), the need for such new physics and potential terms should also be apparent on transforming to the Einstein frame. But no modifications to the model in the Einstein frame are necessary. Thus the breakdown of the perturbation theory calculation in the Jordan frame does not indicate the need for new physics, but simply that perturbation theory in the Jordan frame is not an appropriate method to calculate the smoothly evolving cross-section.

\subsection{Quantum Gravity Effects}

We still have to consider a possible new physics cut-off associated with quantum gravity. Again thinking of particle collisons in the vacuum, we expect quantum gravity effects to become strong at $E \sim  M_{P}$. This it true in both the Jordan and Einstein frames since $\Omega = 1$ in the vacuum, as in the preceeding discussion. Therefore it is possible that new physics exists, characterised by $M_{P}$. In this case terms suppressed by powers of $M_{P}$ should be added to the scalar potential. Since $M_{P}$ is a physical mass, such terms should be introduced in the physical Jordan frame, therefore the scalar potential $V(h)$ will contain terms of the form\footnote{This choice of a new physics cut-off at $M_{P}$ in the Jordan frame corresponds to prescription II of \cite{bms}.} $h^{4+n}/M_{P}^{n}$. This possibility has been considered in \cite{bgs}, where it was shown that such terms do not significantly change the predictions of the model. This is understandable since the value of $h$ during inflation is $h \approx \sqrt{N} M_{P}/\sqrt{\xi} \ll M_{P}$.  We note that the energy scale at which quantum gravity in the Einstein frame becomes strong, $\tilde{E} \sim M_{P}$, where $\tilde{E} \equiv\Omega(h) E$ is the energy with respect to the Einstein frame metric, is then less than the new physics scale in the Einstein frame, $
\Omega(h) M_{P}$, when $h$ is large and $\Omega(h) \gg 1$. However, this does not introduce a problem for Higgs inflation, since during inflation $\tilde{\rho} \approx \lambda M_{P}^{4}/4 \xi^{2} \ll M_{P}^{4}$ and so quantum gravity effects in the Einstein frame are negligible.

\subsection{Einstein frame and $\Lambda$}

    In \cite{barbon}, the question of whether the hypothetical problems associated with the scale $\Lambda$ could manifest themselves directly in the Einstein frame was also addressed. They considered the scalar potential for $\chi$ in the Einstein frame. If we expand \eq{e7} as a series in $\chi$ when
$\xi h /M_{P} \ll 1$ then we obtain (setting $v = 0$)
\be{e25}  U(\chi) \approx \frac{\lambda}{4}\chi^4 - 3 \lambda
\frac{\xi ^2 \chi^6}{M_{P}^2} + ...   ~.\ee
A factor $\Lambda^{-2}$ appears in the $\chi^6$ term. This is what one might expect if $\Lambda$ represented a new scale of physics, and one might then expect the potential to rapidly grow once $\chi \gtrsim \Lambda$, making the inflaton potential unnatural at $\chi \gtrsim \Lambda$. However, this is not the case. We know the complete $U(\chi)$ in the Einstein frame. In fact, the full potential becomes flatter as $\chi$ increases, tending from $\chi^4$ at $\chi \ll \Lambda$ to $\chi^2$ at $\chi \gg \Lambda$. The apparent generation of new physics is an artifact of considering only two terms of the expansion when all terms are important. Therefore there is no new physics occuring at $E$ or $\chi \sim \Lambda$ in the Einstein frame, in contradiction to what one would expect if such new physics were to appear in the Jordan frame.

\subsection{Power counting arguments}

Finally we comment on the discussion of \cite{burgess}. They use the power-counting formalism of effective field theory to  dimensionally estimate the amplitude for Higgs-Higgs scattering via graviton
exchange, including the contribution of loop corrections. They then apply the condition that the cross-section is unitary to obtain an upper bound on the energy $E$,  reaching the same conclusion as
\cite{barbon}, $E \lesssim M_{P}/\xi \equiv \Lambda$. However, this method assumes that perturbation theory is valid at $E \gtrsim \Lambda$. As in the case of \cite{barbon}, since the theory is completely perturbative and unitary in the Einstein frame, the breakdown of unitarity in Higgs scattering must indicate a breakdown of perturbation theory as a calculational method in the Jordan frame, not a fundamental breakdown in the theory itself requiring new physics to be introduced at $E \gtrsim \Lambda$ to restore unitarity.

      Supporting evidence for this view provided by an analysis of Han and Willenbrock (HW), which considered scalar particle scattering via s-channel graviton exchange \cite{hw}. Firstly, they showed that tree-level unitarity violation occurs only once perturbation theory is no longer reliable. Specifically, imaginary 1-loop contribution to the amplitude is 1/2 of the tree-level contribution, at the onset of tree-level unitarity violation. This is consistent with our expectation that it is perturbation theory and not unitarity which breaks down in the Jordan frame. Secondly, HW make a remarkable observation. They consider the large-$N$ limit of the scattering amplitude, where $N$ is the weighted sum of the particles contributing to the 1-loop diagram, while keeping $N/M_{P}^2$ fixed. In this case they are able to sum all the higher-loop contributions to the scattering amplitude, with the result that the full (i.e. non-perturbative) scattering cross-section satisfies unitarity, even though unitarity is violated at tree-level.   This demonstrates that it is possible for unitarity to 
appear to be violated at tree-level and at low orders of perturbation theory, even though the full cross-section is well-behaved. We expect that a similar phenomenon is occuring in the Jordan frame of Higgs-type inflation, with a well-behaved cross-section when calculated non-perturbatively.

\section{Conclusions}

     Higgs inflation is a completely consistent and natural slow-roll inflation model when studied in the Einstein frame, with no breakdown of perturbation theory or unitarity in scattering processes and no new physics implied by such processes. Based on this, and on the Equivalence Theorem of the S-matrix under non-linear field transformations, which implies that the Higgs-Higgs scattering rate in the Jordan and Einstein frames are the same, we conclude that the apparent breakdown of the calculation of Higgs-Higgs scattering via s-channel graviton exchange in the Jordan frame at $E \gtrsim \Lambda$ does not signify a new physics scale. It signifies only the failure of the calculational method, perturbation theory. In particular, Higgs-Higgs scattering in the Jordan frame does not imply new terms in the Higgs potential scaled by powers of $\Lambda$, since the need for such new physics and potential terms would also be apparent in the Einstein frame. Put simply, since the Higgs inflation model is a completely consistent (perturbative and unitary) model in the Einstein frame, it must be a completely consistent model in the Jordan frame.

 Although we have focused on the Higgs inflation model, the same conclusions apply to any model with inflation due to a non-minimally coupled scalar field, in particular the $S$-inflation model of \cite{rl}.

        In our discussion we addressed the issue of the naturalness with respect to perturbativity and unitarity violation.  However, one might question the naturalness of Higgs inflation in more general terms. In particular, a non-minimal coupling $\xi \simeq 10^4$ might seem unnaturally large. But inflation models often rely on extremely small scalar couplings, $\lambda \simeq 10^{-13}$ - nine orders of magnitude further away from 1. In this sense, Higgs inflation could be considered a great improvement in terms of naturalness.

\section*{Addendum} 

    After this paper was written, a new paper appeared \cite{comm1} which also considers the naturalness of Higgs inflation from the point of view of the Jordan and Einstein frames. (For a related comment on our paper, see \cite{comm2}.) This paper confirms our results in the case where only a single real Higgs scalar and its self-interactions are included (as is assumed in our analysis and in previous discussions \cite{barbon, burgess}), namely that there is no breakdown of unitarity at the energy scale $\Lambda$. This also disproves the previous claims of \cite{barbon} and \cite{burgess} that new physics is necessary in the single scalar model (a point not made clear in \cite{comm2}). However, \cite{comm1} also argues that inclusion of additional non-minimally coupled scalars (in particular, the Goldstone bosons of the Higgs doublet) would make it impossible to simultaneously redefine the scalar kinetic terms for all scalars to minimal form, and that dimension 6 interactions scaled by $\Lambda$ will then cause unitarity violation in the Einstein frame. (A similar point is made in \cite{comm2}.) This results in the remarkable conclusion that inflation based on a single non-minimally coupled scalar conserves unitarity and requires no new physics but with two or more non-minimally coupled scalars the theory violates unitarity and breaks down at $E \sim \Lambda$. A pion analogy is suggested in \cite{comm1} which may explain the suppression of unitarity violation in the single scalar case. Although the arguments given in \cite{comm1} are clear and convincing on first reading, we believe a critical assessment of these non-trivial results will be necessary before they can be accepted as definitive. This is particularly true in light of the previously widely-accepted and superficially convincing but ultimately incorrect claims of \cite{barbon} and \cite{burgess} that the single scalar model breaks down at $E \sim \Lambda$.  We hope to report on such a study in the future.

\section*{Acknowledgements}

We would like to thank Qaisar Shafi for correspondence. 

\vspace{0.1cm}

This work was supported by the European Union through the Marie Curie Research and Training Network "UniverseNet" (MRTN-CT-2006-035863).

\section*{Appendix: Relating the number of e-foldings in the Jordan and Einstein frames}

\renewcommand{\theequation}{A-\arabic{equation}}
 \setcounter{equation}{0}

The definition of the number of e-foldings $\tilde{N}$ appearing in the slow-roll expression for the spectral index in the Einstein frame differs from the physical number of e-foldings $N$ appearing in the Jordan frame.
This is because the definitions of the scale factor in the Einstein and Jordan frames are different. In the Einstein frame
\be{e14} \tilde{N} = \ln \left( \frac{\tilde{a}_{end}}{\tilde{a}} \right)
= \ln \left( \frac{a_{end}}{a}  \frac{\Omega(t_{end})}{\Omega(t)} \right) = N + \ln\left( \frac{\Omega(t_{end})}{\Omega(t)} \right) ~,\ee
where $a_{end}$ and $t_{end}$ are the scale factor and time at the end of inflation.
During inflation $\Omega^{2}(t) \approx \xi h^2/M_{P}^2$. Therefore $\Omega(t_{end})/\Omega(t) \approx h(t_{end})/h(t)$.
Solving the slow-roll equation for $\chi$
\be{e17} 3 H \dot{\chi} = -\frac{ \partial V}{\partial \chi} \approx - \frac{\lambda M_{P}^{3}}{\sqrt{6} \xi^2} e^{-\frac{2 \chi}{\sqrt{6} M_{P}}}  ~,\ee
implies that
\be{e18} \tilde{N} = \frac{3}{4} \frac{\xi}{M_{p}^{2}} \left[h^2 - h_{end}^2\right] \approx \frac{3}{4} \frac{\xi}{M_{p}^{2}} h^2 \;\;\;,\;\; h^2 \gg h_{end}^{2} ~,\ee
where $h$ is related to  $\chi$ via \eq{e10}.
Inflation ends when $\tilde{\eta} \approx 1$, which implies $h_{end}^{2} \approx 4 M_{p}^{2}/3 \xi $.
Therefore $\Omega(t_{end})/\Omega(t) \approx 1/\sqrt{\tilde{N}}$ and
\be{e20} \tilde{N} \approx N + \ln \left( \frac{1}{\sqrt{\tilde{N}}} \right)    ~.\ee
This difference is important when calculating the spectral index corresponding to a given length scale. In principle, the number of e-foldings $N$ for a given length scale can be known precisely in Higgs inflation, because the reheating temperature is precisely determined by Standard Model interactions. The value of the spectral index in Higgs inflation from the slow-roll approximation is (for the classical scalar potential)
\be{a1} n = 1 - \frac{2}{\tilde{N}} - \frac{3}{2 \tilde{N}^2} + O\left(\frac{1}{\tilde{N}^3} \right)    ~.\ee
This gives a value $n = 0.9663$ at $\tilde{N} = 60$. However, if we evaluate it at $N = 60$, which corresponds to
$\tilde{N} = 57.95$, then we obtain $n = 0.9650$. Since the value of $n$ for the classical potential is a key parameter of Higgs inflation, the shift due to the definition of $N$ must be correctly taken into account.


\end{document}